\documentclass[12pt]{article}
\usepackage{cite}
\usepackage{amsmath, amsthm, amssymb,slashed}
\usepackage{ifpdf}
\ifpdf
  \usepackage[pdftex]{graphicx}
  \usepackage{epstopdf}
\else
  \usepackage[dvips]{graphicx}
\fi
\parskip=\baselineskip
\def\Bbb{\mathbb}
\def\Tr{{\rm Tr}}
\def\16{{\bf 16}}
\def\1{{\bf 1}}
\def\2{{\bf 2}}
\def\3{{\bf 3}}
\def\4{{\bf 4}}
\def\i{{\mathrm i}}

\def\R{{\Bbb{R}}}\def\Z{{\Bbb{Z}}}
\def\N{{\mathcal N}}

\font\teneurm=eurm10 \font\seveneurm=eurm7 \font\fiveeurm=eurm5
\newfam\eurmfam
\textfont\eurmfam=\teneurm \scriptfont\eurmfam=\seveneurm
\scriptscriptfont\eurmfam=\fiveeurm

\font\teneusm=eusm10 \font\seveneusm=eusm7 \font\fiveeusm=eusm5
\newfam\eusmfam
\textfont\eusmfam=\teneusm \scriptfont\eusmfam=\seveneusm
\scriptscriptfont\eusmfam=\fiveeusm

\font\tencmmib=cmmib10 \skewchar\tencmmib='177
\font\sevencmmib=cmmib7 \skewchar\sevencmmib='177
\font\fivecmmib=cmmib5 \skewchar\fivecmmib='177
\newfam\cmmibfam
\textfont\cmmibfam=\tencmmib \scriptfont\cmmibfam=\sevencmmib
\scriptscriptfont\cmmibfam=\fivecmmib

\numberwithin{equation}{section}

\def\d{\mathrm d}

\def\Z{{\Bbb Z}}

\def\t{\widetilde}

\def\S{{\sf S}}
\def\SU{\mathrm{SU}(2)}
\def\be{\begin{equation}}
\def\ee{\end{equation}}
\def\SO{\mathrm{SO}}
\def\T{{\mathcal T}}
\begin{document}
\begin{titlepage}
\begin{flushright}
\end{flushright}

\vskip 1in
\begin{center}
{\bf\Large{A Note On Some Minimally Supersymmetric Models}\vskip.3cm{ In Two Dimensions}}
\vskip
1cm\centerline {Davide Gaiotto and Theo Johnson-Freyd}\vskip .05in
{\small\textit{Perimeter Institute, 31 Caroline St N, Waterloo, ON N2L 2Y5, Canada}} \vskip 1cm
\centerline{and}
\vskip
0.5cm {Edward Witten} \vskip 0.05in {\small{ \textit{School of
Natural Sciences, Institute for Advanced Study}\vskip -.4cm
{\textit{Einstein Drive, Princeton, NJ 08540 USA}}}
}
\end{center}
\vskip 0.5in
\centerline{\bf{Abstract}}
\vskip.05in{We explore the dynamics of a simple class of two-dimensional models with $(0,1)$ supersymmetry, namely sigma-models with target $\S^3$ and the minimal possible set of fields.
For any nonzero value of the Wess--Zumino coupling $k$, we describe a superconformal fixed point to which we conjecture that the model flows in the infrared.  For $k=0$, we conjecture
that the model spontaneously breaks supersymmetry.   We further explore the question of whether this model can be continuously connected to one that spontaneously breaks
supersymmetry by ``flowing up and down the renormalization group trajectories,'' in a sense that we describe.    We show that this is possible if $k$ is a multiple of 24, or equivalently if the
target space with its $B$-field is the boundary of a ``string manifold.''    The mathematical theory of ``topological modular forms'' suggests that this condition is necessary as well as sufficient. }
\baselineskip 16pt
\date{September, 2011}
\end{titlepage}

\tableofcontents
\section{Introduction}

This paper is devoted to some aspects of supersymmetric models in two spacetime dimensions with the minimal possible supersymmetry, often called $(0,1)$ supersymmetry,
that is, one supersymmetry for right-moving excitations and none for left-moving ones.

In section \ref{specific}, we consider the dynamics of a very  specific class of  model with $(0,1)$ supersymmetry.
We consider a sigma-model with target space a three-sphere $\S^3$.  We assume the minimal field content, so the only degrees of freedom are the sigma-model
fields and their superpartners.   Apart from a coupling constant that represents the radius of the sphere, the theory has an integer-valued Wess--Zumino coupling $k$.
For $k=0$, the  model has a global symmetry O(4), rotating the sphere; for $k\not=0$, this is reduced to SO(4).    

The model is asymptotically-free at short distances, and the
question is to determine its behavior at long distances.   For sufficiently large $|k|$, the model flows to an infrared fixed point that can be described reliably at weak coupling.
Extrapolating this to small $|k|$, we propose, for any nonzero $k$, a specific superconformal fixed point to which we conjecture that the model flows in the infrared.   The model is
a simple supersymmetric WZW model, with a somewhat subtle action of SO(4).    The basis for the conjecture is that the fixed point that we propose is correct for large enough $k$
and reproduces  't Hooft anomalies of the SO(4) symmetry for any $k$.    For $k=0$, no reasonable candidate superconformal fixed point presents itself and we conjecture that
the model spontaneously breaks supersymmetry.    

In section \ref{flowing}, we pose the following question: can the model be perturbed to trigger supersymmetry breaking?   This question requires some clarification.   First of all,
we do not limit ourselves to perturbations that preserve the SO(4) symmetry; the only global symmetry that a perturbation is supposed to maintain is the (0,1) super Poincar\'{e} symmetry.   In that framework, the obvious notion of perturbing the model is to add a marginal or relevant operator at the microscopic level.   Then one flows to the infrared
and asks if supersymmetry is spontaneously broken.   However, we want to consider a more general notion of perturbation, in which, roughly speaking, one is allowed to flow up as well as down
the renormalization group (RG) trajectories.   
Let $\T$ be some  $(0,1)$  supersymmetric theory.    We allow ourselves to add arbitrary massive degrees of freedom
to theory $\T$ in a supersymmetric fashion.   In more detail, we allow ourselves to
 replace theory $\T$ with any other theory $\T'$ that is equivalent to $\T$ at long distances.    Then we  perturb
theory $\T'$ in an arbitrary (supersymmetric) fashion, to get some other theory $\T''$, and ask whether theory $\T''$ breaks supersymmetry at low energies.    We refer to  this process
$\T\to \T'\to \T''$ as ``flowing up and then down the RG trajectories.''    More generally, we permit ourselves to flow up and then down repeatedly.   The 
question
is whether in this sense a given theory $\T$ can be continuously connected, via a family of supersymmetric theories, to a theory that spontaneously breaks supersymmetry.

By an explicit construction, we show in section \ref{flows}  that the sigma-model with target $\S^3$
  can be continuously connected in this sense to one that spontaneously breaks supersymmetry
provided that $k$ is divisible by 24.

As explained in section \ref{tmf}, there is reason to believe that this condition is necessary as well as sufficient.     There is a mathematical theory of ``topological modular
forms'' (TMF) \cite{Hopkins,TMF} that is closely related to $(0,1)$ supersymmetric models in two dimensions.   In TMF theory, one studies  invariants of a familiar type --- basically
the elliptic genus \cite{LS}, which can be interpreted as
a supersymmetric index  \cite{Witten} --- but one also defines more subtle torsion invariants.   Building in part  on earlier work \cite{Se}, it has been proposed that every (0,1) theory defines a class in TMF \cite{ST,ST2}, and this class is supposed
to depend, in some sense, only on the homotopy class of the (0,1) theory.    This conjecture motivated
a recent analysis of certain holomorphic SCFT's \cite{GJF}, and was also recently
discussed in the context of a construction of two-dimensional (0,1) models by compactification from six dimensions \cite{Vafa}.  A physical interpretation or consequence of
the conjecture appears to be that if the TMF class associated to a given model is nonzero, then the model  is not continuously connected --- even by flowing repeatedly up and down the RG trajectories ---
to one that spontaneously breaks supersymmetry.   
The TMF class of a sigma-model with target $\S^3$ is simply the value of $k$ mod~24.  So  the conjecture means that these models cannot be continuously
connected to one that spontaneously breaks supersymmetry if $k$ is not divisible by 24.    No immediately obvious physical argument  would explain this.
One possible solution will be proposed 
 elsewhere \cite{TJ}.

\section{Some $(0,1)$ Models And A Conjecture Concerning Their Dynamics}\label{specific}

To describe a sigma-model with target $\S^3$, we introduce scalar fields $X_I$, $I=1,\cdots,4$, with a constraint
\be\label{woggo} \sum_{I=1}^4 X_I^2=1.  \ee
The basic sigma-model action on a two-manifold $\Sigma=\R^2$ is
\be\label{noggo} \frac{1}{2\lambda}\int_\Sigma \d^2x \sum_I\partial_\mu X_I \partial^\mu X_I. \ee
To this, one might add a Wess--Zumino coupling, as discussed shortly.

The sigma-model without the Wess--Zumino coupling has an obvious O(4) symmetry.   We will usually add such a coupling, explicitly breaking O(4) to SO(4).    A double
cover of SO(4) is a product of two SU(2) groups, say $\SU_\ell\times\SU_r$, and we will express most statements in terms of these $\SU$'s.
The disconnected component of O(4) exchanges the two SU(2)'s.    

Now let us add a set of four right-moving\footnote{We also refer to right- or left-moving fermions as  fermions of positive or negative chirality. Strictly speaking the field
 has positive or negative chirality, while the corresponding excitation is right- or left-moving.}
 fermions $\psi_I$ that transform in the vector representation of O(4) (or SO(4)), just like the $X_I$.    To say that the fermions are right-moving
means that if we introduce light cone coordinates $u=t-x$, $v=t+x$ (where the Lorentz signature metric is $\d s^2=-\d t^2+\d x^2$), then the action for the $\psi_I$ is
\be\label{toggo}\frac{\i}{2}\int \d^2 x \sum_I \psi_I \frac{\partial}{\partial v} \psi_I. \ee

Because these fermions are purely right-moving, they have anomalies under the global symmetry.  These anomalies would obstruct gauging the global symmetry, but more to the point for
our purposes, as in \cite{thooft}, they constrain the possible behavior of the theory in the infrared.   Anomalies for a simple nonabelian Lie group in two dimensions are quantized
as integer multiples of a basic invariant.   Fermions in the vector representation of $\SO(4)\sim\SU_\ell\times \SU_r$ have the 
smallest possible nonzero anomaly.   We will describe this by saying that the anomaly of this theory under $\SU_\ell\times \SU_r$ is $(1,1)$ (in units of the basic anomaly of an $\SU$ group).

An important point is that in two dimensions, the sign of the contribution of a given mode to a global symmetry anomaly depends only on whether the mode is right-moving or left-moving.
    Thus the $\psi_I$ make contributions of the same sign to the  $\SU_\ell$ and $\SU_r$ anomalies.
This is consistent, of course, with the fact that a symmetry in the disconnected component of O(4) exchanges the two subgroups.   Our convention will be that right-movers make positive contributions
to anomalies and left-movers make negative contributions.

Now we want to remove one of the fermions, by imposing a constraint
\be\label{nogogo}\sum_{I=1}^4 X_I\psi_I=0. \ee
The model constrained in this way actually has $(0,1)$ supersymmetry.   It will be described in section \ref{flowing} in a manifestly supersymmetric way.   For now, let us just
note that the fields of a supersymmetric sigma-model are bosonic fields $X$ that describe a map $X:\Sigma\to M$  of spacetime $\Sigma$ to some target space $M$, and fermion fields
which are spinors on $\Sigma$ valued in $X^*(\mathrm TM)$, the pullback of the tangent bundle of $M$. In the case of a $(0,1)$ model, the fermions are more specifically positive chirality fermions on $\Sigma$ valued in $X^*(\mathrm TM)$.    In the present case, $M=\S^3$, and the constraint (\ref{nogogo}) is the right one to ensure that the $\psi_I$
are valued in the pullback of the tangent bundle to $\S^3$. 
   A (1,1) sigma-model in two dimensions has non-minimal four-fermi couplings, but there are no such couplings in a (0,1) model that has only right-moving
fermions, so the constraint (\ref{nogogo}) is enough to give a (0,1) supersymmetric model.

Imposing this constraint removes one ($X$-dependent) component of $\psi_I$, but it preserves the $\SU_\ell\times \SU_r$ symmetry, and does not change the anomalies, which
remain $(1,1)$.   To see that the anomalies are unchanged, we can proceed as follows.   Introduce a left-moving  SO(4)-singlet fermion $\chi$, with action
\be\label{ploggo}\frac{\i}{2}\int\d^2x \,\chi \frac{\partial}{\partial u}\chi. \ee
As $\chi$ is an SO(4) singlet, adding it to the theory does not affect the anomalies.    Now add a mass term
\be\label{woggox}\i m \int \d^2x \, \chi\sum_{I=1}^4 X_I\psi_I. \ee
Since this term can be turned on continuously, it does not affect the $\SU_\ell\times \SU_r$ anomalies of the theory.   
With $m\not=0$, both $\chi$ and one ($X$-dependent) component of $\psi$ becomes massive and disappear from the low energy theory.   At low energies the constraint
(\ref{nogogo}) emerges, since the components of $\psi$ that remain massless are the ones that satisfy this constraint.    One may describe this by saying that at energies low compared to $m$,
the kinetic energy (\ref{ploggo}) of $\chi$ becomes unimportant, and $\chi$ behaves as a Lagrange multiplier imposing the constraint.

This procedure of obtaining the supersymmetric model by perturbing a model in which $\psi$ is unconstrained can be done in a manifestly supersymmetric way, as we will explain in section 
\ref{noapp}.   For now let us simply note that even though the procedure was not manifestly supersymmetric, at low energies the model automatically becomes supersymmetric.   What ensures
this is that the supersymmetric model with the constraint (\ref{noggo}) has no marginal or relevant deformation that preserves its O(4) symmetry.   

Now let us add a Wess--Zumino interaction, with a coefficient $k$.   
We first briefly address the purely bosonic sigma-model with target $\S^3$.
The Wess--Zumino interaction 
contributes $(-k,k)$ to the $\SU_\ell\times \SU_r$ anomalies.
(The overall sign is a convention; what is important is that the two groups receive anomalies of opposite signs.)    
As
 analyzed in \cite{WittenTwo}, if $|k|$ is
sufficiently large, one can explicitly find in perturbation theory a weakly coupled fixed point, known as the WZW model at level $k$.   
This fixed point has left-moving and right-moving current algebra symmetries $\SU_L$ and
$\SU_R$, both at level $|k|$.   (By unitarity, current algebra levels are nonnegative.)    The values of these levels means that $\SU_R$ has an anomaly $|k|$ and $\SU_L$ has an anomaly
$-|k|$.   The minus sign just reflects the fact that the $\SU_L$ current algebra is left-moving.

There is a subtlety here, which is that at short distances both $\SU_\ell$ and $\SU_r$ couple to both left- and right-moving degrees of freedom, but in the infrared one of them couples
only to left-moving degrees of freedom and one only to right-moving degrees of freedom.
  To find which is which, we compare the anomalies.   Since the anomalies of $\SU_\ell\times \SU_r$ are $(-k,k)$ and the anomalies of
$\SU_L\times \SU_R$ are $(-|k|,|k|)$, we have     $\SU_\ell=\SU_L$, $\SU_r=\SU_R$ if $k>0$, and the opposite relationship $\SU_\ell=\SU_R$, $\SU_r=\SU_L$ if $k<0$.   

From the standpoint of weakly coupled perturbation theory, this picture is only reliable for sufficiently large $|k|$.   However, it is believed that  this picture is valid for all $k$.   
In the special case $k=0$,
the current algebras become trivial and the theory is gapped.

Now let us discuss the (0,1) supersymmetric version of the model.  In general, adding  fermions does not affect the renormalization group equations of a two-dimensional sigma model
 in one-loop order \cite{AF}.   Therefore, for sufficiently
large $|k|$, the model flows to a weakly coupled fixed point, known as the $(0,1)$ supersymmetric WZW model.
  At this fixed point, there is a left-moving current algebra $\SU_L$ at some level $\kappa$, and a right-moving
$\N=1$ supersymmetric $\SU_R$ current algebra, also at level $\kappa$.   
Modular invariance of the combined theory implies that the two levels are equal. 
The bosonic analysis implies that
  $\kappa$ is equal to $|k|$ plus
a correction of order 1 due to the fermions.   We will determine this correction in a moment.   We have necessarily $\kappa\geq 0$, since this is required for unitarity.

A supersymmetric $\SU_R$ current algebra at level $\kappa$ actually has an $\SU_R$ anomaly $\kappa+2$.   In fact the supersymmetric current algebra can be constructed from an ordinary
bosonic $\SU_R $ current algebra at level $\kappa$, which contributes $\kappa$ to the anomaly, together with  three chiral fermions in the adjoint representation of $\SU_R$, which contribute 2  to the anomaly.\footnote{In more detail,
the super current algebra for a simple Lie group $G$ is generated by dimension 1/2 superfields ${\mathcal J}^a(u,\theta)=\eta^a(u)+\theta J^a(u)$ in the adjoint representation of $G$.
Having dimension 1/2, the $\eta^a$ are free fermions, as discussed in related examples in \cite{OG}. The $J^a$ are currents of dimension 1.
  For the super current algebra to have level $\kappa$, the  $J^a$
 are a sum of level $\kappa$
currents $\t J^a$ that commute with the $\eta^a$ and a bilinear $f^a_{bc}\eta^b\eta^c$, where $f^a_{bc}$ are the structure constants of $G$. 
The fermionic currents $f^a_{bc}\eta^b\eta^c$ generate a current algebra of level $h^\vee$ (the dual Coxeter number of $G$).  So a level $\kappa$ representation of the super current algebra,
when viewed as a representation of the ordinary current algebra of $G$, has level $k=\kappa+h^\vee$.  
The ordinary current algebra is what measures
 the anomaly in the $G$ symmetry.
  For $G=\SU$, $h^\vee=2$ and $k=\kappa+2$. }  
On the other hand, the ordinary $\SU_L$ current algebra at level $\kappa$ has an anomaly $-\kappa$.      In the special case $\kappa=0$, the $\SU_L$ current algebra is trivial and the $\SU_R$ current algebra is realized entirely by three free fermions in the adjoint representation of $\SU_R$.

Another way to 
say
this is that the (0,1) supersymmetric WZW model at 
 level $\kappa$  is equivalent to an ordinary WZW model also at level $\kappa$
plus three free right-moving
fermions in the adjoint representation of $\SU_R$, contributing 2 to the $\SU_R$ anomaly.   The (1,1) supersymmetric WZW model at level $\kappa$ was originally discussed in \cite{AA,K} and
this behavior was found.    The same arguments apply in the (0,1) case; to get from (1,1) to (0,1), one just sets the left-moving fermions to zero in the classical action.  

In short, in the ultraviolet, we have $\SU_\ell\times \SU_r$ symmetry with anomalies $(-k+1,k+1)$, and in the infrared with have $\SU_L\times \SU_R$ symmetry with anomalies $(-\kappa,\kappa+2)$,
where moreover $\kappa\geq 0$.   Comparing these formulas, we find the relation between $k$ and $\kappa$ and the relation between $\SU_\ell\times \SU_r$ and $\SU_L\times \SU_R$.
If $k>0$, then $\kappa=k-1$ and $\SU_\ell=\SU_L$, $\SU_r=\SU_R$.    If $k<0$, then $\kappa=-k-1$ and $\SU_\ell=\SU_R$, $\SU_r=\SU_L$.  

Thus, for any nonzero $k$, there is a natural candidate for a superconformal field theory to describe the infrared limit of the $(0,1)$ sigma-model with target $\S^3$, and we conjecture
that the model does flow to this fixed point.    For $k=0$, there is no evident candidate fixed point, since $\kappa$ would have to be negative,
 and we conjecture that the model spontaneously breaks supersymmetry.  For $k=\pm 1$,
the candidate fixed point consists just of three right-moving free fermions, transforming in the adjoint representation of $\SU_r$ or of $\SU_\ell$, depending on the sign of $k$.  For $|k|>1$,
there is in addition an SU(2) WZW model at level $|k|-1$.   

This proposal is reminiscent of the behavior of $k$ coincident NS5-branes in Type II superstring theory.  (Some statements in this paragraph will be phrased assuming that $k\geq 0$;
if $k<0$, one considers antibranes instead of branes.)   For a system of coincident NS5-branes, string perturbation theory can break down in a ``throat'' region where the
effective string coupling constant becomes large.   The throat region is believed to be described by an $\S^3$ sigma-model with (1,1) supersymmetry together with a super Liouville field.   String perturbation theory breaks down because the effective string coupling constant diverges at the end of the throat.
This breakdown of perturbation theory is believed to occur precisely for $|k|\geq 2$.  (A single fivebrane is believed to be described by a well-behaved and throatless superconformal field theory, though not
one that is known in any explicit form; with no fivebranes at all, one simply has the superconformal field theory of $\R^4$.)   The assertion that perturbation theory breaks down only for
$|k|\geq 2$ matches nicely with the fact that $(1,1)$ superconformal field theories that are candidates as infrared limits of the $(1,1)$ sigma-model of $\S^3$ exist precisely if $|k|\geq 2$
\cite{Seiberg}.   In our problem, since we are considering $(0,1)$ supersymmetry, the analog is to consider coincident heterotic string fivebranes \cite{CHS}, which  can arise
as small $\mathrm{Spin}(32)/\Z_2$ or ${\mathrm E}_8\times {\mathrm E}_8$ instantons.    It is believed that in this case, string perturbation theory breaks down even in the
presence of a single small instanton \cite{Wittensmall}.   So a superconformal field theory of $\S^3$ which (together with a super Liouville field) can describe a throat region is needed for all
$|k|\geq 1$, and this is what we have found.

Another example with somewhat similar behavior is a gauge theory in three spacetime dimensions with a simple nonabelian group $G$, a Chern-Simons level $k$,
and minimal supersymmetry (which in 
three dimensions
means two supercharges).   For  $|k|\geq h^\vee/2$, where $h^\vee$ is the dual Coxeter number of $G$, the theory is believed to 
flow to a weakly coupled gapped ground state, leading to a topological field theory (TFT) that can be described explicitly 
in terms of a purely bosonic Chern-Simons theory
.
But
 for $|k|<h^\vee/2$ it is believed that supersymmetry is spontaneously broken \cite{WittenThree}. For $G=\SU$, $h^\vee=2$ and the exceptional case with supersymmetry breaking occurs only for $k=0$.
We will return to this analogy in section \ref{interpolation}.

For $k=0$, but not otherwise, the $(0,1)$ model with target $\S^3$ can be generalized to a model with target $\S^N$.  For large $N$, the model can be studied in a $1/N$ expansion
by standard methods, and one can show that supersymmetry is spontaneously broken.  This is briefly sketched at the end of section \ref{noapp}.
 Thus our conjecture about the $k=0$ case amounts to claiming that this behavior persists down to $N=3$.
\def\X{{\mathcal X}}
\def\g{\h g}
\def\h{\widehat}
\def\tr{{\mathrm{tr}}}

\section{Flowing Up and Down the RG Trajectories}\label{flowing}

\subsection{Review Of $(0,1)$ Models In Superspace}\label{reviewsuper}

For completeness, we will first review a few standard facts about $(0,1)$ models
in superspace.   We give only a brief explanation of the main facts we will use.  See \cite{Gates,Gates2} for early work.

We work in a superspace $\h\Sigma$ with bosonic light cone coordinates $u,v$ and a fermionic coordinate~$\theta$.  Lorentz
boosts act by $u\to e^w u$, $v\to e^{-w}v$,
$\theta\to e^{w/2}\theta$.   In $(0,1)$ supersymmetry, there is a single fermionic symmetry.   It is generated by 
\be\label{nono}Q=\frac{\partial}{\partial\theta}+\i \theta\frac{\partial}{\partial u} \ee
and (anti)commutes with the superspace derivative
\be\label{bono}D=\frac{\partial}{\partial\theta}-\i \theta\frac{\partial}{\partial u} ,\ee
along with the ordinary derivatives $\partial_u$ and $\partial_v$.    

For our purposes, we will consider two kinds of superfield.    To describe a map $\X:\h\Sigma \to M$, where $M$ will be the target space of a sigma-model, we use scalar superfields
$\X^I(u,v,\theta). $
Geometrically, $\X^I$ represents the pullback of a system of local coordinates on $M$ to $\h\Sigma$, via a map $\X:\h\Sigma\to M$. 
We can expand
\be\label{wono}\X^I(u,v,\theta)=X^I(u,v)+\i\theta\psi^I(u,v), \ee
where $X^I$ are bosonic fields describing a map $X:\Sigma\to M$ ($\Sigma$ is the reduced space of $\h\Sigma$, parametrized by $u$ and $v$), and the $\psi^I$ are chiral fermion fields
on $\Sigma$ valued in the pullback to $\Sigma$ of the tangent bundle to $M$.
  A nonlinear sigma-model with target $M$
can then be described at the classical level by the action
\be\label{lono}I_\X=\frac{\i}{2} \int \d u \d v \d\theta\,\,\g_{IJ}(\X)\partial_v \X^I D\X^J, \ee
where $\g_{IJ}$ is a completely arbitrary second rank tensor on $M$.   
This action is manifestly supersymmetric, since the derivatives used, namely $\partial_v$ and~$D$, commute with the supersymmetry generator~$Q$.
We can expand $\g_{IJ} $ as a sum of a symmetric part~$g_{IJ}$ and an antisymmetric part~$B_{IJ}$:
\be\label{zono} \g_{IJ}=g_{IJ}+B_{IJ}. \ee
The symmetric part is a metric tensor on $M$, and the antisymmetric tensor corresponds to a two-form $B$ on $M$.  In string theory, $B$ is called the $B$-field.

For example, if we take $B=0$, then integrating over $\theta$ gives a particularly simple action in terms of ordinary fields
\be\label{rono}I_\X= \frac{1}{2}\int \d u \d v\left(g_{IJ}\partial_u X^I \partial_v X^J +\i g_{IJ} D_v\psi^I \psi^J\right) . \ee
For the special case that $M=\S^3$ with a round metric, this is essentially the model that was discussed in section \ref{specific}, after imposing the constraint (\ref{noggo}).

Now let us discuss the role of the $B$-field.   At the classical level, a short computation reveals that the part of the action that is proportional to $B$ is invariant under shifting $B$ by an exact form,
that is under 
\be\label{untu} B\to B+\d\Upsilon, \ee
where $\Upsilon$ is a 1-form on $M$.    What is invariant under this operation is a three-form, the ``field strength''
\be\label{nutr}H=\d B. \ee
The Bianchi identity at the classical level says that
\be\label{nonx}\d H=0. \ee
The model can make sense even if    $H$ cannot  be written as $\d B$ with a globally defined $B$.  However, $H$ has to obey a condition of ``Dirac quantization,'' which says that $H/2\pi$ has integer periods
(on any three-cycle on $M$).   A more precise statement, taking torsion and discrete topological effects into account, is that to define the model, one has to choose a class 
$x\in \mathrm{H}^3(M,\Z)$ that can be represented modulo torsion by the three-form $H/2\pi$.    For our example of $M=\S^3$, we have $H^3(M,\Z)\cong \Z$, so the choice of $x$ is just
the choice of the integer $k$ that appeared in section \ref{specific}:
\be\label{wonx} k=\int_{S^3}\frac{H}{2\pi}. \ee  If the characteristic class $x$ of the $B$-field is non-zero, then the $B$-field cannot really be represented globally as a two-form and the form 
(\ref{rono}) of the action is only valid locally.

At the quantum level, one runs into certain anomalies in quantizing this model.   First of all, one finds that $M$ has to be a spin manifold.  More subtle is the ``sigma-model anomaly,'' on which
an early reference is \cite{MN}; it plays an important role in the Green--Schwarz mechanism of anomaly cancellation in heterotic string theory.  The upshot of the sigma-model anomaly is the following.
Any Riemannian manifold $M$ has a first Pontryagin class $p_1\in \mathrm{H}^4(M,\Z).$   If, however, $M$ is a spin manifold, then there is a characteristic class $\lambda$ with the property that
$2\lambda=p_1$; it is best to write formulas in terms of $\lambda$ rather than $p_1$, because (as there may be two-torsion in $\mathrm{H}^4(M,\Z)$) one may lose information in multiplying by 2.  
In terms of differential forms, $\lambda$ is represented by $-(1/16\pi^2)\tr\, R\wedge R$, where $R$ is the curvature two-form of $M$, regarded as a matrix acting on the tangent bundle of $M$
(where the trace is taken).   The sigma-model anomaly says that rather than $H$ being a closed three-form (as is the case classically), it is a trivialization of $\lambda$.   
In topology, a choice of (spin structure and) trivialization of $\lambda$ is called a ``string structure'' on $M$.

In terms of differential
forms, this means that
 $\d H$ is not zero at the quantum level, but rather
\be\label{doofus}\d H =-\frac{1}{8\pi} \tr\, R\wedge R. \ee 
In particular,
 although in general $H$ cannot be interpreted as a closed three-form, it does make sense to shift $H$ by a closed three-form.   The assertion that the $B$-field trivializes $\lambda$
as an integral cohomology class, and not just at the level of differential forms, depends on an analysis of global anomalies.   See \cite{Wittenfour}, section 2, for an explanation of this
and a more precise explanation of the meaning
of the $B$-field at the quantum level.

We will also need another type of superfield, called 
a fermi superfield.   For our purposes, a fermi superfield is an anticommuting superfield $\Lambda$ that transforms as a negative chirality spinor.   It has an expansion
\be\label{nonxip}\Lambda(u,v,\theta)=\chi+\theta F, \ee
where $\chi$ is a negative chirality fermion field on $\Sigma$ and $F$ is a scalar auxiliary field.  (There is a generalization in which the fermi superfields are valued in the pullback $\X^*(V)$
of some real vector bundle  $E\to M$, but we will only consider the case that $E$ is trivial and rank 1.)   The kinetic energy for this field is
\be\label{onxip}I_\Lambda=\frac{1}{2}\int\d u \d v \d\theta\, \Lambda D\Lambda =\frac{1}{2}\int \d u \d v\left(\i\chi \partial_u\chi +F^2\right). \ee
Thus $\chi$ is a massless fermion of negative chirality, and $F$ is a scalar ``auxiliary field,'' that is, the action does not depend on its derivatives.
We can also couple a fermi superfield to scalar superfields as follows.   If $W$ is any real-valued function on $M$,  called a superpotential, we can add to the action a term
\be\label{wonxip}I_W=\int\d u \d v \d\theta\, \Lambda W(\X^I)=\int\d u\d v\,\left( F W(X^I)+\i\chi \sum_I\psi_I \frac{\partial W(X^K)}{\partial X^I}\right). \ee
After ``integrating out'' the auxiliary field $F$ (or in other words eliminating it by solving its equation of motion), we can replace the sum $I_\Lambda+I_W$ with
\be\label{nxip}I'=\int\d u \d v\left( \frac{\i}{2}\chi\partial_u\chi -\frac{1}{2}W^2 -\i\chi \sum_I\psi^I \frac{\partial W(X^K)}{\partial X^I}\right). \ee
In particular,  there is a potential energy
\be\label{toffo} V=\frac{1}{2}W^2. \ee
In addition, at any point in field space with $\d W\not=0$, $\chi$ and a linear combination of the $\psi^I$ combine to get a mass.

\subsection{Application to the $\S^3$ Model}\label{noapp}

Now we want to describe a manifestly supersymmetric linear sigma-model realization of the model with target $\S^3$ that was described in section \ref{specific}, for the special case $k=0$.
We introduce four scalar superfields $X^I$ that transform in the vector representation of O(4).   The $X^I$ parametrize $M=\R^4$, with a flat metric $g_{IJ}=\delta_{IJ}$.   We also introduce
a single fermi superfield $\Lambda$.   We couple $\Lambda$ to the $X^I$ with the superpotential
\be\label{noffo} W= \sum_I X_I^2-R^2, \ee
where $R$ will be the radius of the sphere.   The potential energy of the model is just
\be\label{noffox} V(X)=\frac{1}{2}\left(\sum_I X_I^2-R^2\right)^2. \ee 
At low energies, $X$ will be confined to a sphere of radius $R$.   In addition, the Yukawa coupling $\chi\psi_I \partial_IW$ in eqn.~(\ref{wonxip}) becomes $2\chi \sum_I X_I\psi_I$, which is the
coupling that was assumed in eqn.~(\ref{woggox}).   Thus the model is equivalent  at low energies (after obvious rescalings and identifications of parameters) to the $k=0$ model as studied in section \ref{specific}.   

We can also generalize to the case that $X^I$ is an $N+1$-component field with O($N+1$) symmetry.  The low energy limit is then a sigma-model with target $\S^N$.   This model can be analyzed
in the large $N$ limit using standard methods; see for example \cite{Alv, WittenFive}, where similar models with more supersymmetry were analyzed.   
As is typical in such calculations, the scalar superfields $X^I$ acquire a mass for large $N$.  What is
special, however, to the case of $(0,1)$ supersymmetry, in contrast to models with more supersymmetry, is that since the mass is generated from the coupling
\be\label{wonfo} \int \d u\d v\d\theta\, \Lambda \left(\sum_I X_I^2-R^2\right), \ee
the mechanism for the $X_I$ to get a mass is that an auxiliary field gets an expectation value, namely the  auxiliary field $F$ in the multiplet $\Lambda=\chi+\theta F$.   This spontaneously breaks supersymmetry, with 
$\chi$ as the Goldstone fermion.   Spontaneous supersymmetry breaking does not  occur for large $N$ in analogous models with more supersymmetry.   

  We note that the generalization of the $\S^3$ sigma-model to include the Wess--Zumino coupling $k$ does not have an analog for $\S^N$ with $N>3$.  That is because $\mathrm{H}^3(\S^N,\Z)=0$ when $N\not=3$.   

\subsection{Flowing Up And Down}\label{flows}

\def\Y{{\mathcal Y}}
We want to explore whether it is possible to perturb the $\S^3$ sigma-model at level $k$ so as to break supersymmetry.    What we mean by ``perturbing'' the model was explained in the
introduction:   we are allowed to first replace the model with any other model that flows in the infrared to the $\S^3$ sigma-model, and then to make a conventional perturbation of that new
model.   In the introduction, we described this as a process of flowing up and down the RG trajectories.   

The Wess--Zumino coupling $k$, because of its topological nature, is not easily seen in a linear sigma-model, at least at the classical level.   
Hence we will study the $\S^3$ model in this section as a nonlinear sigma-model, without trying to derive it from a linear model.   

So we include 4 scalar superfields $\X_I$ with a constraint  $\sum_I \X_I^2=R^2$.   The constraint means that these scalar superfields parameterize a sphere $\S^3$ of radius $R$.
Taking the standard metric on the sphere, the action $I_\X$ of eqn. (\ref{rono}) is the desired nonlinear sigma-model action. 

Now we will add massive degree of freedom to the model in a supersymmetric fashion.  We do this first in a way that does not really give anything new, but that will be a useful
starting point.   Ultimately, we will add massive degrees of freedom in a way that preserves supersymmetry but not the O(4) or SO(4) symmetry of the model.

  We replace the target $\S^3$ of
the sigma-model with $\R\times \S^3$, adding a new scalar superfield $\Y=Y+\i\theta\zeta$ that describes a map from $\h\Sigma$ to $\R$.   We also add a fermi multiplet $\Lambda=
\chi+\theta F$.   For the action of these fields, we take
\be\label{pingo} I_{\Y,\Lambda}=\int\d u\d v\d\theta\left( \frac{\i}{2}\partial_v \Y D\Y +\frac{1}{2}\Lambda D\Lambda +m\Lambda(\Y-y_0)\right) ,\ee
for some real constants $y_0$ and 
$m$
(the latter is  
a mass parameter). Thus the superpotential is $W=m(\Y-y_0)$.   After integrating out the auxiliary field $F$, the potential energy is $\frac{1}{2}m^2(Y-y_0)^2$.   So the field $Y$ gets an expectation value $y_0$ and becomes massive.
Similarly the fermionic components of $\Lambda$ and $Y$ combine to gain a mass $m$.  All of the added degrees of freedom are massive, so at low energies, we recover  the $\S^3$ nonlinear sigma-model that we started with.

To get something interesting, let $Z$ be any four-dimensional compact spin manifold.   $Z$ must be a spin manifold to avoid an anomaly in the $(0,1)$ sigma-model with target $Z$. 
Even then, the $(0,1)$ sigma-model with target $Z$ is anomalous if $\int_Z\lambda\not=0$.     This will be the interesting case in what follows.   To get
a well-defined $(0,1)$ model, we cannot then take the target space to be $Z$.   But let us remove a point from $Z$ to get a noncompact four-manifold $Z'$.
    Removing
a point makes the $\lambda$ class topologically trivial, since there is now no compact four-cycle on which it could be integrated.  By projecting the missing point to infinity, one
can define a complete Riemannian metric on $Z'$.  $Z'$ has a noncompact end that is topologically $\R\times \S^3$,   and we choose the metric on $Z'$ to be asymptotically tubular, 
which is to say
we assume that the metric on $Z'$ looks
asymptotically like  the product of a round metric on
${\S^3} $ times 
a flat metric on
$\R$.
    Thus, in the asymptotic portion of field space, the model consists of a free chiral superfield parametrizing~$\R$, and decoupled
from it a nonlinear sigma-model with target $\S^3$.   Importantly, this sigma-model may have a Wess--Zumino coupling.   This is because the anomaly equation (\ref{doofus}) implies
that
\be\label{yfo}\int_{\S^3}\frac{H}{2\pi} = -\int_{Z'} \frac{\tr\,R\wedge R}{16\pi^2}.  \ee
But with the assumption that we have made about the metric of $Z'$,
the  integral on the right hand side of eqn. (\ref{yfo}) has a simple topological meaning.   We could ``cap off'' the noncompact metric on $Z'$, giving us back $Z$, by adding a hemisphere at the
end.   If we give the cap a standard round metric, then $\tr\,R\wedge R=0$ in the cap, and hence the integral on the right hand side of eqn. (\ref{yfo}) can be replaced by an integral on $Z$.
This integral on $Z$ is the topological invariant $\int_Z\lambda$.    So in other words
the Wess--Zumino coupling on $\S^3$ is the integral of the $\lambda$ class over $Z$:
\be\label{nygo}\int_{\S^3}\frac{H}{2\pi}=\int_Z\lambda.\ee

For a four-dimensional compact spin manifold $Z$, the invariant $\int_Z\lambda$ is always an integer multiple of 24.    To prove this, note that according to the Atiyah-Singer index theorem,
the index $\mathcal I$ of the Dirac operator on $Z$ is ${\mathcal I}=\int_Z p_1/24=\int_Z\lambda/12$.  On the other hand, because the spinor representation of $\mathrm{Spin}(4)$ is pseudoreal, 
$\mathcal I$ is always even.   So $\int_Z\lambda$ is divisible by 24.   This is the only constraint, since if $Z$ is a K3 surface, then $\int_Z\lambda = 24$.  Moreover,  a connected sum of K3 surfaces,
possibly with reversed orientation, can give any integer multiple of 24.   

The model as described so far has {\it massless} degrees of freedom beyond the $\S^3$ sigma-model, so it is not what we need.    To proceed, add a fermi multiplet $\Lambda$ to the theory,
and couple it to the nonlinear sigma-model with target  $Z'$ with a superpotential coupling
\be\label{wondo}I_W=\int \d u \d v \d\theta\, \Lambda W, \ee
where $W$ is a  function on $Z'$. 

To describe what sort of function we want, parametrize $\R$ by a real variable $Y$ and let us suppose that $Z'$ coincides with (or can be well approximated by) $\R\times\S^3$ for $Y>0$,
while the part of $\R\times \S^3$ with $Y\ll0$ is ``missing,'' being capped off in some fashion by $Z'$.   Then, for some constant $y_0>0$, we can find a function $W$ on $Z'$ that
coincides with $m(Y-y_0)$ for $Y>0$, and is negative-definite outside of the region $Y>0$.  Of course, any such function is bounded below, since the complement  in $Z'$ of the region $Y>0$
is compact.
  The coupling  $I_W$ introduces a potential energy $\frac{1}{2}W^2$, and this vanishes only on $\{y_0\}\times \S^3$.
As in our earlier discussion of the model with target $\R\times \S^3$, the low energy theory is just a sigma-model with target $\S^3$; other degrees of freedom are massive.
  The Wess--Zumino coupling $k$ in this sigma-model
is $\int_Z\lambda$, according to eqn. (\ref{nygo}), and thus can be any integer multiple of 24.

Now we can introduce a perturbation of the model that will trigger supersymmetry breaking.   We replace the function $W$ with $\t W=W+c$, where $c$ is a positive constant.  Since $W$ is bounded
below, we see that if $c$ is sufficiently large, then $\t W$ is positive-definite.  In this case, the potential energy, which now is $\frac{1}{2}\t W^2$, is everywhere strictly positive, so supersymmetry
is spontaneously broken.  

We have learned that, if $k$ is divisible by 24, then  the (0,1) sigma-model with target $\S^3$ and Wess--Zumino coupling $k$ can be continuously connected to a model with spontaneously
broken symmetry by, loosely speaking,
 flowing up and down the RG trajectories.   We flowed up the RG trajectories by replacing $\S^3$ with $Z'$ 
equipped with
 the superpotential $W$.   Then we made an ordinary
perturbation from $W$ to $\t W$, and flowed down the RG trajectories to find that this model spontaneously breaks supersymmetry.

The construction that we have discussed is related as follows to small instantons of the heterotic string (this subject was briefly discussed at the end of section \ref{specific}).   First, 
 consider a (0,1) sigma-model with target space a smooth K3 surface $Z$.
To cancel the anomaly, we can add $n$ left-moving worldsheet fermions coupled to an SO($n$) bundle $E\to Z$ of instanton number 24.   In the limit that the instantons become small, the
left-moving fermions decouple and the instantons are replaced by heterotic string fivebranes, originally analyzed in  \cite{CHS}.    Such a fivebrane is described by omitting from $Z$ a point $p$ 
(the position of the fivebrane) and placing on the complement of $p$ a tubelike metric, asymptotic to $\R\times \S^3$, with $p$ understood to lie at infinity in this description.   In this
description, the $B$-field flux on $\S^3$ satisfies $\int_{\S^3}H/2\pi = n$, where $n$ is the number of small instantons that have collapsed to the point $p$.   Our construction amounts
to the special case that all 24 instantons collapse to the same point $p$.    Superconformal symmetry (as opposed to global (0,1) supersymmetry) was not important in our discussion.
However, if one starts with a superconformal field theory of a smooth K3 surface with an SO($n$) bundle chosen to cancel the anomaly, then in the small instanton limit, 
the dilaton field will grow linearly in the $\R$ direction.   With such an asymptotically linear dilaton field, the sigma-model with target $Z'$ is in fact  superconformal.

\subsection{String Cobordism}\label{sc}

We have emphasized the example of $M=\S^3$, but actually what we have just explained has an analog for any compact string manifold $M$.
We mentioned string structures already in section~\ref{reviewsuper}: a manifold $M$ is ``string''  or is a string manifold if it is equipped with  a spin structure and also a trivialization of the class $\lambda = p_1/2 \in \mathrm{H}^4(M,\Z)$.
   We say that $M$ is the string boundary of $Z$
if $Z$ is a string manifold of one dimension more such that $M$ is the boundary of $Z$ in the usual sense, and the string structure of $Z$ restricts on its boundary to the string structure of $M$.
In this situation, we can make precisely the construction that was just described.   Let $Z'$ be the open manifold obtained from $Z$ by omitting its boundary.   On $Z'$, we can place a complete
Riemannian metric that near infinity looks like $\R\times M$.   Then we can make the same construction as in section \ref{flows}, introducing the same sort of superpotential as before
and deforming it in the same way.

From this construction we
 learn 
that if $M$ is a string boundary, then the (0,1) sigma-model with target~$M$ can be deformed, by flowing up and down, to one that spontaneously breaks supersymmetry.

\subsection{A Three-Dimensional Interpolation}\label{interpolation}

Now we will return to the problem studied in section \ref{specific}.   However, we will consider a $(0,1)$ sigma-model with target an arbitrary compact simple Lie group $G$,
not necessarily $\SU$.   Moreover, instead of discussing renormalization group flows, we will consider a more general type of supersymmetric interpolation between two theories.

We will show how to interpolate supersymmetrically between a $(0,1)$ sigma-model with target space  $G$ 
and Wess--Zumino coupling $k$ and a  simple supersymmetric WZW model.

Consider a three-dimensional gauge theory with $N=1$ supersymmetry, gauge group $G$, Yang-Mills coupling $g$ and Chern-Simons level $k_{3d}$.   The fields are a gauge field $A$
and an adjoint-valued Majorana fermion $\chi$.  
Place the theory on a product geometry $ {\mathbb R}^2\times [0,L]$, where the second factor is parametrized by a coordinate $x_3$.	
Impose Dirichlet boundary conditions $A=0$ at both ends.   It is not possible to extend this
boundary condition
 to $\chi$ in a way that preserves all of the supersymmetry, but
a simple boundary condition $\gamma_3\chi|=\chi|$ at both ends preserves a $(0,1)$ supersymmetry in the two-dimensional sense.   (A boundary condition
$\gamma_3\chi|=-\chi|$ at both ends would preserve a $(0,1)$ supersymmetry of the opposite chirality, but if the signs are opposite at the two ends, all supersymmetry
is broken.)   

With Dirichlet boundary conditions on gauge fields, one requires the generator of a gauge transformation to vanish on the boundary, so there is no gauge anomaly.   However, 
Dirichlet boundary conditions support a boundary $G$ global symmetry, acting by a gauge transformation  that goes to a constant at the boundary.   Thus, there is a global
symmetry $G$ at each end.   
This global symmetry has an 't Hooft anomaly which receives a contribution  both from the bulk Chern-Simons level and from the chiral boundary condition for 
 $\chi$.  The contribution to the anomaly from the Chern-Simons term is $k_{3d}$ at one end and $-k_{3d}$ at the other.    The fermions make a contribution
$\frac{h^\vee}2$ at each end.%
\footnote{We could reverse this sign by reversing the sign in the  boundary condition for $\chi$, but the sign of the fermion  anomaly is
the same at each end because $\chi$ obeys a boundary condition with the same sign at each end.   If we set $g=0$, $\chi$ has a zero-mode in the $[0,L]$ direction
that describes at low energies a massless chiral two-dimensional fermion in the adjoint representation of $G$.   Such a fermion has an anomaly $h^\vee$.   In the present context,
there is no anomaly in bulk, and half of the usual anomaly lives at each end, somewhat  as in \cite{HW}.}
Thus the anomaly coefficient is $\frac{h^\vee}{2}+ k_{3d}$ for the $G$ global symmetry at one end, and $\frac{h^\vee}{2}-k_{3d}$ for the $G$ global symmetry at the other end. 

The system on the slab will thus have a $G_\ell \times G_r$ global symmetry, with 't Hooft anomaly coefficients $\frac{h^\vee}{2} \pm k_{3d}$.   Thus the sum of these coefficients is $h^\vee$,
and this is the anomaly under a diagonal subgroup of $G_\ell\times G_r$.    

The simplest observable charged under the boundary global symmetries is an open Wilson line stretched between the two boundaries. 
If $g^2 L \ll 1$, then the Dirichlet boundary conditions kill most of the gauge degrees of freedom except for the holonomy across the $[0,L]$ segment. 
That holonomy is $G$-valued, and the system is well approximated by a $(0,1)$ sigma model with target $G$ and Wess--Zumino coupling $k = k_{3d}$.

If $g^2 L \gg 1$, though, we need to understand the three-dimensional gauge dynamics in the bulk and near each boundary. As noted in section \ref{specific}, 
the three-dimensional gauge theory with $N=1$ supersymmetry is expected to break SUSY if $|k_{3d}|<\frac{h^\vee}{2}$. If $|k_{3d}| \geq \frac{h^\vee}{2}$, 
it will flow to a 3d topological field theory (TFT) which is essentially equivalent to the one which arises from a bosonic $G$ Chern-Simons theory at level $k' = k_{3d} - \frac{h^\vee}{2} \, \mathrm{sign}\, k_{3d}$.

As the bulk theory flows to a TFT, the boundary should flow to a relative\footnote{A relative 2d CFT obeys axioms similar to those of an ordinary CFT, except that it is defined on a conformal
two-manifold $\Sigma$ that is the boundary of an oriented three-manifold $M$, which is endowed with some TFT.}
 two-dimensional CFT which has the correct global symmetries and 't Hooft anomalies. 
 We will focus on the case $|k_{3d}| \geq \frac{h^\vee}{2}$. Then it is very natural, and fully consistent, for the IR boundary theory to be either a left-moving 
chiral WZW model at level $\pm k'$ or a supersymmetric right-moving chiral WZW model at level $\pm k'$, depending on the sign of $k'$ and of the orientation of the boundary. 
These are the simplest 2d theories which are compatible with 
supersymmetry, boundary 't Hooft anomalies, and being relative to the $G$ Chern-Simons TFT  at level $k'$.

If this is so, then the slab system for $g^2 L \gg 1$ flows in the IR to the  simple supersymmetric WZW model that we discussed (for $G=\SU$) in section \ref{specific}.

\section{A Puzzle}\label{tmf}

There is 
in mathematics
a (generalized)
 cohomology theory called ``topological modular forms'' (TMF) \cite{Hopkins,TMF}
 in which an ``orientation'' of a manifold $Z$ is a choice
of string structure, i.e.\ 
a spin structure together with a $B$-field that trivializes the $\lambda = p_1/2$ class. 
(An ``orientation'' in a generalized cohomology theory is the data needed to ``integrate'' a cohomology class over a manifold. For instance, K-theory is ``spin-oriented'' because there is a natural way to integrate a K-theory class over a spin manifold, due to the relationship between K-theory and Dirac operators.)

As reviewed in section~\ref{reviewsuper},
a string structure is the topological information that is needed to define an anomaly-free (0,1) sigma-model with target $Z$.   Such a model has integer-valued topological
invariants that are relatively well-known: they appear in the elliptic genus \cite{LS,Witten}.   The variant of the elliptic genus that is relevant here
is obtained by compactifying the model on a circle and then defining in the Ramond sector  the supersymmetric index $Z(q)=\Tr\,(-1)^F q^{L_0}$, where $L_0=(H-P)/2$ is the energy
of left-moving modes ($H$ and $P$ are the Hamiltonian and the momentum).  A path integral representation shows that $Z(q)$ is a modular function; the coefficients
in its $q$-expansion are integers, because of their interpretation as an index.   

In TMF theory, however, one also defines torsion invariants of a manifold $Z$ with string structure.  The physical meaning of these torsion invariants is less apparent.

It has been proposed \cite{ST,ST2} that every (0,1) theory in two dimensions defines a class in TMF.   This class is supposed to be invariant under deformations
of the (0,1) theory.   Although there is certainly not a complete understanding, it would appear that the class of allowed deformations should be at least large enough to allow
the operations that we have considered: 
 flowing  up and down the RG trajectories, while making conventional deformations along the way.

The physical meaning of the conjecture would appear to be that a (0,1) model that represents a nonzero class in TMF cannot be deformed to a model that spontaneously breaks supersymmetry,
since such a model should define the zero class in TMF.

In the special case of a sigma-model with three-dimensional compact oriented target space $M$, 
the class $\lambda \in \mathrm{H}^4(M,\Z)$ vanishes for dimensional reasons, and so a string structure consists of a choice of spin structure together with
any choice of a $B$-field.  
The class of the model in TMF  is invariant under shifting $\int_M H/2\pi$ by a multiple of 24.


For  $M=\S^3$, what we have called $k=0$ is the case that the model actually has the full $\mathrm{O}(4)$ symmetry including the orientation-reversing component.  Its TMF class then vanishes (and we conjectured
in section \ref{specific} that it spontaneously breaks supersymmetry without any perturbation). In general,
the TMF class of the model is simply the value of $k$ mod 24.   So TMF theory seems to suggest that the model cannot be deformed to one that breaks supersymmetry unless $k$ is a multiple of
24.

It is not obvious  from a physical point of view why this would be true; if it is correct, it means that there are obstructions to supersymmetry breaking that are not yet appreciated. 
This question will be further explored elsewhere \cite{TJ}.   Here we make only a negative observation.
   The difference between $k$ divisible by 24 and not divisible by 24 cannot be detected by compactifying the model on a circle 
 and asking whether supersymmetry
is spontaneously broken (in the Ramond sector) after this compactification.  We can see why this is true by considering the problem for either large or small radius.

  If the radius $\rho$ of the circle is small, then we can analyze the low energy states on a circle in the sigma-model language.   States
with energy much less than $1/\rho$ can be studied by ignoring modes that carry momentum around the circle.    In the approximation of ignoring those modes, the supercharge reduces,
independent of $k$, to the Dirac operator on $\S^3$, acting on sections of the spin bundle of $\S^3$.  This operator has no zero-modes, so in this limit, supersymmetry is spontaneously broken, regardless of $k$.   On the other hand, if the conjecture
of section \ref{specific} concerning  long distance behavior  is correct, then it can be used to analyze the behavior for large $\rho$.   For $k=0$, the conjecture says that supersymmetry is broken
in the large volume limit and therefore also for sufficiently large $\rho$.   For $k\not=0$, assuming the conjecture, the question amounts to whether the supersymmetric current algebra
of $G$ at level $\kappa=|k|-1$ has a supersymmetric state in the Ramond sector, in other words a state of strictly zero energy.    The answer to this question is that the ground state
energy of the supersymmetric current algebra in the Ramond sector is strictly positive.\footnote{The bosonic current algebra has central charge $c=3\kappa/(\kappa+2)$, leading to a ground
state energy $-c/24=-\kappa/8(\kappa+2)$.  On the other hand, three free fermions in the Ramond sector have ground state energy $+3/24=1/8$.   The sum is then $1/4(\kappa+2)$, and is strictly positive.}
Thus, given the conjecture, supersymmetry is spontaneously broken for large $\rho$ after compactifying on a circle, regardless of $k$.

If there is a reasonable physical observable that detects the TMF class of a theory in this family,  
then obviously it must be sensitive to the $B$-field.  Moreover, assuming the conjecture of section \ref{specific},
this observable should take the same value whether evaluated in the sigma-model or in the corresponding supersymmetric WZW model.

\vskip1cm
\noindent{\it Acknowledgments}
Research of EW is supported in part by NSF Grant PHY-1606531. Research at Perimeter Institute is supported by the Government of Canada through Industry Canada and by the Province of Ontario through the Ministry of Research \& Innovation.
\bibliographystyle{unsrt}

\end{document}